\documentclass[nohyper,11pt,letterpaper]{JHEP3}
\usepackage[dvips]{epsfig}

\usepackage{epsf,amsfonts,amssymb}

\newcommand{\beq}{\begin{equation}}
\newcommand{\eeq}{\end{equation}}
\newcommand{\ben}{\begin{displaymath}}
\newcommand{\een}{\end{displaymath}}
\newcommand{\bea}{\begin{eqnarray}}
\newcommand{\eea}{\end{eqnarray}}
\newcommand{\bean}{\begin{eqnarray*}}
\newcommand{\eean}{\end{eqnarray*}}
\newcommand{\nn}{\nonumber \\}
\newcommand{\ba}{\begin{array}}
\newcommand{\ea}{\end{array}}
\newcommand{\bi}{\begin{itemize}}
\newcommand{\ei}{\end{itemize}}

\def\L {\Lambda}
\def\a {\alpha}
\def\b {\beta}
\def\g {\gamma}
\def\G {\Gamma}
\def\d {\delta}
\def\D {\Delta}
\def\s {\sigma}
\def\e {\epsilon}

\def\r {\rho}

\def\G{\Gamma}
\def\g{\gamma}
\def\e{\epsilon}
\def\m{\mu}
\def\n{\nu}
\def\p{\pi}

\newcommand{\tr}{\mbox{Tr}}

\title{\LARGE Holographic Renormalization of 
Probe D-Branes in AdS/CFT}

\author{Andreas Karch,$^{a}$ Andy O'Bannon$^{a}$
  and Kostas Skenderis$^{b}$ \\
  $^a$ Department of Physics, University of Washington, Seattle, WA 98195-1560 \\
  $^b$ Institute for Theoretical Physics, University of Amsterdam, Valckenierstraat 65, 1018 XE Amsterdam, The Netherlands\\

E-mail: \email{karch@phys.washington.edu, ahob@u.washington.edu, skenderi@science.uva.nl}}

\abstract{We perform holographic
renormalization for probe branes in $AdS_5 \times S^5$. We show that
for four known probe D-branes wrapping
an $AdS_m \times S^n$, the counterterms needed to render the
action finite are identical to those for the free,
massive scalar in $AdS_m$
plus counterterms for the renormalization of the volume of $AdS_m$.
The four cases we consider are the probe D7, two different probe D5's
and a probe D3. In the D7 case there are scheme-dependent finite counterterms 
that can be fixed by supersymmetry. }


\preprint{ITFA-2005-50\\
hep-th/0512125}

\begin{document}

{\vskip 1cm}

\section{Introduction} \label{intro}

Holography equates a $(d+1)$-dimensional theory of gravity with a
$d$-dimensional field theory. The AdS/CFT correspondence
\cite{Maldacena:1997re} is an explicit example of this, in which
supergravity (SUGRA) formulated on the ten-dimensional bulk spacetime $AdS_5
\times S^5$ is dual to $\mathcal{N}=4$ $SU(N)$ super Yang-Mills (SYM)
theory living on the four-dimensional boundary of the $AdS_5$ factor. The
precise statement of the duality equates the on-shell bulk action with
the generating functional of the CFT \cite{Witten:1998qj, Gubser:1998bc}. In general, however, both of these quantities are infinite: the bulk action suffers IR divergences while UV divergences
appear on the CFT side. The method of holographic renormalization
\cite{Henningson:1998gx, Henningson:1998ey,
Balasubramanian:1999re, deHaro:2000xn} was developed to remove
these divergences in a consistent way so that the correspondence equates
finite, physical quantities.

Holographic renormalization involves regulating the on-shell bulk
action and then adding covariant counterterms to cancel the
divergences that appear as the regulator is removed. This technique
is very general, and should certainly apply to the subset of bulk
solutions that describe supersymmetric (SUSY) probe brane solutions in
AdS/CFT. The AdS/CFT duality is constructed from the near-horizon
limit of a stack of $N$ D3-branes with $N \rightarrow \infty$.
Introducing a finite number of additional branes orthogonal to the
D3's in some directions will produce probe branes embedded in the
near-horizon geometry of the D3's, which is $AdS_5 \times S^5$. The simplest
known probe branes of this type preserve half the supersymmetry and
introduce additional fields in the $\mathcal{N} = 4$ SYM. Their
action is the Dirac-Born-Infeld (DBI) action, whose contribution to
the bulk action is simply the volume of the brane. Since the
backreaction of the probe branes on the background geometry can be
neglected this system is probably the simplest application of
holographic renormalization in a string theory setting and serves as
a nice illustration of the method. One aspect of holographic
renormalization for submanifolds, the conformal anomaly, was
previously studied in \cite{Graham:1999pm}. In this paper, we
consider four known probes: a D7-brane
\cite{Karch:2002sh}, two D5-branes \cite{Karch:2000gx,
Skenderis:2002vf}, and a probe D3-brane
\cite{Constable:2002xt}. One of the cases in non-supersymmteric 
\cite{Skenderis:2002vf} while the other three cases are
half  supersymmetric. The brane embedding in all of these cases
is described by a single scalar field that, via the correspondence,
will be dual to some operator in the field theory. 

Given a renormalized bulk action, renormalized correlators in the dual field theory can be computed straightforwardly. The goal of this paper is to compute renormalized one-point functions, or vacuum
expectation values (vev's), for the operators dual to the embedding
scalars, using holographic renormalization. Especially for the D3/D7 system the situation is interesting since
the vev is not just the subleading term in the asymptotic expansion
of the scalar field. Our calculation of the vev clarifies the somewhat ad hoc procedure that
has been used in the literature.

This paper is organized as follows. In section \ref{holorg}, we
review holographic renormalization, including the methods of adding
covariant counterterms to render the 
bulk action finite and of computing one-point
functions from this renormalized action. In section \ref{general}, we show
that the counterterms for our DBI actions are identical to the
counterterms for a free scalar in $AdS$. In section \ref{d7}, for the D7,
we use these counterterms to compute the one-point function for the
embedding scalar's dual operator. In this case, we find that a finite
counterterm must be included if the renormalization scheme is to be
supersymmetric. We do a similar computation of a one-point function in
section \ref{d5} for the D5 probe. We also consider a second D5 embedding that is not considered in the general arguments of section \ref{general} but which needs no new counterterms. In Section \ref{d3} we compute one-
and two-point functions for the D3 probe.

\section{Review: Holographic Renormalization} \label{holorg}

We review in this section the holographic renormalization of the volume of an
$AdS_{d+1}$ spacetime and of a free scalar in that spacetime. The former
has been studied extensively in \cite{Henningson:1998gx,
Henningson:1998ey,
Balasubramanian:1999re, Graham:1999jg}, 
while the latter was explained in detail in
\cite{deHaro:2000xn}, see \cite{Skenderis:2002wp} for a review of
holographic renormalization.

To begin, write the metric of asymptotically $AdS_{d+1}$ spacetime in the
form of Fefferman and Graham \cite{Fefferman} in units where the AdS
radius is one (which we use throughout):
\beq
ds^{2} = G_{\mu \nu} dx^{\mu} dx^{\nu} = \frac{dr^2}{r^2} + \frac{1}{r^2}g_{ij}(x,r)dx^i dx^j,
\label{metric}
\eeq
where r is the radial coordinate, $\mu, \nu = 0, \ldots, d$ and $i,j = 0,
\ldots, d-1$ and where
\beq
g(x,r) = g_{(0)} + r^2 g_{(2)} + \ldots + r^d g_{(d)} + h_{(d)} \log(r^2) + O(r^{d+1}).
\eeq
The logarithmic term only appears for even $d$ and only even powers
of $r$ appear up to order $r^{d-1}$. Solving Einstein's equation recursively 
for the $g_{(i)}$ gives two useful identities\footnote{Indices are raised
and lowered using $g_{(0)}$.}: 
$\tr g_{(2)} = \frac{R_0}{2(d-1)}$, with $R_0$ the Ricci scalar built from 
$g_{(0)}$, and 
$\tr g_{(4)} 
= \frac{1}{4} \tr (g^2_{(2)})$ 
\cite{deHaro:2000xn}.

For the renormalization of the volume of $AdS_{d+1}$, we restrict our discussion to $d=2,3,4$, which will be relevant for the probe branes we consider. 
The volume is\footnote{This is related with the on-shell value of the
Einstein-Hilbert action for gravity with a negative cosmological constant
as,
$V_{AdS} = \frac{\kappa^2}{d} S_{AdS}^{on{-}shell}$,
where 
$S_{AdS} = \frac{1}{2 \kappa^2} \int d^{d+1} x \sqrt{G} (R + 2 \L).$
Notice that we use the conventions in \cite{deHaro:2000xn}, in particular
the curvature conventions are such that AdS has positive curvature.}
\beq
V_{AdS} = \int d^{d+1}x \sqrt{G}\, .
\label{ads}
\eeq
 \noindent Naively, this volume is infinite since the integration extends
all the way to the boundary at $r=0$. Holographic renormalization proceeds
in two steps. First, a regulator is introduced by extending the
integration only to $r = \e$. Second, counterterms are added to cancel $\e
\rightarrow 0$ divergences, yielding a finite, physical answer. These
counterterms must be built from data on the $r = \e$ slice to preserve
covariance. This will include the induced metric on the $r=\e$ slice,
$\g_{ij}$, and the Ricci scalar (and tensor) built from $\g$, $R_{\g}$.
This counterterm procedure is valid for any on-shell bulk action, which
will in general be divergent due to integration down to $r=0$, and applies
to both the $AdS_{d+1}$ volume and the free scalar action as we will
review below. The counterterms for the volume renormalization are
\bea
L_{1} &=& - \frac{1}{d} \sqrt{\g}, \label{ct1} \nn 
L_{2}  &=& \left \{
\begin{array}{cccl}
        && - \frac{1}{2} \frac{1}{d(d-1)(d-2)} \sqrt{\g} R_{\g}
 & \, \, \, \,\mbox{ for }d \, \neq \, 2 \\
        && \frac{1}{4} \log(\e) \sqrt{\g} R_{\g}
 & \, \, \, \,\mbox{ for }d \, = \, 2
\end{array}
\right .
\label{ct2}
\eea
and for $d=4$ only, an additional counterterm is needed:
\begin{eqnarray}
L_{3} & = & 
 - \log{\e} \sqrt{\g} \frac{1}{32} 
(R_{ij}R^{ij} - \frac{1}{3} R_{\g}^2).
\label{ct3}
\end{eqnarray}

\noindent We will denote subtracted and renormalized quantities as $V_{sub} = V_{AdS} + \int d^d x [ L_1 + L_2 + L_3 ]$ and $\displaystyle V_{ren} = \lim_{\e \to 0} V_{sub}$.

A free, massive scalar field $\Phi(x,r)$ in this $AdS$ background has
action
\beq
S = \frac{1}{2} \int d^{d+1}x
\sqrt{G} (G^{\mu \nu} \partial_{\mu} \Phi \partial_{\nu} \Phi + M^2 \Phi^2)
\eeq
\noindent with equation of motion (EOM) the Klein-Gordon equation with
the Laplacian associated with $G_{\mu \nu}$. 
Via the usual AdS/CFT dictionary, this
scalar will be dual to some gauge-invariant CFT operator with dimension
$\Delta$ given by $M^2 = (\Delta - d)\Delta$. A solution to the EOM
resulting from the above action will, in general, have the form
\bea \label{Phi_exp} 
\Phi(x,r) & = & r^{d-\Delta} \phi(x,r) \\ & = &
r^{d-\Delta}(\phi_{(0)} + r \phi_{(1)} + \ldots + r^{2\Delta - d}
\phi_{(2\Delta - d)} + r^{2\Delta - d} \log(r^2) \psi_{(2\Delta - d)}) +
O(r^{\Delta + 1}). \nonumber 
\eea
The logarithmic term appears only when $2\Delta - d$ is an integer,
which we will assume to be the case, and reflects the existence of 
matter conformal anomalies in the dual CFT, i.e. conformal anomalies 
due to short distance singularities in correlators of composite 
gauge invariant operators \cite{Petkou:1999fv}. All coefficients $\phi_{(n)}$
for $n < 2\Delta - d$, as well as $\psi_{(2\Delta - d)}$, can be
computed recursively by inserting this form of the solution into the
EOM and expanding in powers of $r$. The result is that the leading
coefficient, $\phi_{(0)}$, and the coefficient $\phi_{(2\D - d)}$
are not fixed by the EOM but all other coefficients are fixed by
these two. The one coefficient we will need explicitly in terms of
$\phi_{(0)}$ is the coefficient of the logarithmic term for the case
$2\D - d = 2$:
\beq
\psi_{(2)} = - \frac{1}{2} [ \Box_0 \phi_{(0)} + (\frac{d}{2} - 1)\phi_{(0)} \tr g_{(2)} ].
\label{psi2}
\eeq

The action naively evaluated on such a solution is again infinite because
of IR divergences, so holographic renormalization proceeds as before.
The resulting counterterms are given by
\bea
L_{4} &=& \sqrt{\g} \frac{(d-\Delta)}{2} ( 1 +
\frac{1}{d-\D} \frac{1}{\log{\e}}) \Phi^{2}(x,\e) \label{ct4} \nn
L_{5} &=& \frac{1}{2(2\Delta - d - 2)} \sqrt{\g}
(\Phi(x,\e)\Box_{\g}\Phi(x,\e) +
\frac{d-\Delta}{2(d-1)}R[\g]\Phi(x,\e)^2) \label{ct5} 
\eea
These counterterms are the only ones required when $2\D - d \leq 2$.
The second term in $L_4$, proportional to $\frac{1}{\log(\e)}$, only
appears in our examples when the scalar saturates the Breitenlohner-Freedman (BF)
bound for $AdS_{d+1}$, $M^2 \geq -\frac{d^2}{4}$ \cite{Bianchi:2001de}, for which $2\D - d = 0$. In this case also
$L_4$ alone suffices to renormalize the action. In the case that $2\D -d = 2$, the coefficient $[2(2\Delta - d - 2)]^{-1}$ in $L_{5}$ is replaced with $-\frac{1}{2} \log(\e)$ \cite{deHaro:2000xn}.

Renormalized correlators in the boundary theory can be computed from the
renormalized action. According to the AdS/CFT dictionary, the leading
coefficient of the asymptotic expansion will act as a source for the dual
operator. Denoting this operator as $O$, the one-point function can be
written
\beq 
\langle O \rangle = \frac{1}{\sqrt{g_{(0)}}} \frac{\d S_{ren}}{\d \phi_{(0)}},
\eeq
\noindent which can in turn be written in terms of quantities on the $r=\e$ slice:
\beq 
\langle O \rangle = \lim_{\e \rightarrow 0} \left(\frac{1}{\e^{\D}}
\frac{1}{\sqrt{\g}}  \frac{\d S_{sub}}{\d \Phi(x,\e)}\right). 
\eeq 
For the cases we consider, if the scalar
saturates the BF bound this will be modified: $\frac{1}{\e^\D}
\rightarrow \frac{\log(\e)}{\e^\D}$ \cite{Bianchi:2001de}.

The structure of the counterterms and the renormalized one-point functions
is most transparent in the radial Hamiltonian formalism, 
where the radius is playing the role of time \cite{Papadimitriou:2004ap,PS2}.
This is also the fastest way to arrive at the results reviewed above,
since in this method covariance is manifest at all stages.
In this formalism the central object is the radial canonical momentum,
\beq
\pi=\frac{\partial L}{\partial \dot{\Phi}}\, ,
\eeq
where overdot denotes differentiation w.r.t. the radius $\rho=- \log r$.
It can be easily shown to correspond to a
regularized one-point function,
\beq \label{reg}
\langle O \rangle_{reg} 
\equiv \frac{\delta S_{reg}}{\delta \Phi(x,\e)} = \pi(x,\e)\, .
\eeq
This expression is infinite as the regulator is removed. To extract the 
finite part in a covariant fashion we use the fact that the theory 
possesses a well-defined dilatation operator,
\beq \label{dil}
\d_D = \int d^d x 
\left(2 \g_{ij} \frac{\delta}{\d \g_{ij}} 
+ (\D -d) \Phi \frac{\delta}{\d \Phi} 
\right).
\eeq
Let us expand the canonical momentum in eigenfunctions of $\d_D$,
\beq \label{exp}
\pi = \sqrt{\gamma} (\pi_{(d-\D)} + \cdots + \pi_{(\D)} +
\tilde{\pi}_{(\D)} \log r^2 +\cdots), 
\eeq
where the subscripts indicate dilatation weight of the terms,
\beq
\d_D \pi_{(n)} = - n \pi_{(n)}, \qquad \d_D 
\tilde{\pi}_{(\D)} = - \D \tilde{\pi}_{(\D)}, \qquad
\d_D \pi_{(\D)} = - \D \pi_{(\D)} -2 \tilde{\pi}_{(\D)}.
\eeq
Notice that the normalizable mode $\pi_{(\D)}$ transforms 
anomalously.
This expansion is a covariant analogue of the radial expansion 
(\ref{Phi_exp}). The explicit form of $\pi_{(i)}$ is found by 
inserting (\ref{exp}) in Hamilton's equations and iteratively solving 
them by collecting terms with the same weight. This procedure determines
all $\pi_{(i)}$ except $\pi_{(\D)}$ which is left undetermined, as in the 
discussion below (\ref{Phi_exp}).

The renormalized 1-point function is now simply given by the 
term of weight $\D$, as it should since the dual operator has dimension 
$\D$,
\beq
\langle O \rangle = \pi_{(\D)}.
\eeq
Furthermore, evaluating the regulating action on-shell one 
immediately gets
\beq \label{ac_reg}
S_{reg} = - \int d^d x \sqrt{\gamma} \frac{1}{2} \pi \Phi,
\eeq
and thus the counterterms are (minus) the terms involving 
the momentum eigenfunctions with weight less than $\D$ plus the logarithmic
term involving $\tilde{\pi}_{\D}$. This is a general feature of the counterterm
action for any theory, namely the counterterms are always certain 
linear combinations of momentum eigenfunctions with weight 
less than $\D$ plus logarithmic counterterms obtained by expanding the 
regulated action in eigenfunctions of the dilatation operator and using the 
basic relation (\ref{reg}) with $\d \to \d_D$,
\beq
\d_D S_{reg} = \pi \d_D \Phi + \pi^{ij} \d_D \g_{ij}
\eeq
where we have reinstated the metric dependence.
We refer to \cite{Papadimitriou:2004ap,PS2} for the details and 
further discussion.

\section{Generalities for Probe Branes} \label{general}

\subsection{General Solution and Counterterms}

All the cases we consider in this section 
are SUSY probe D-branes embedded in $AdS_5
\times S^5$. In every case, the probe wraps some $AdS_m \times S^n$
with $|m - n| = 2$ as required by supersymmetry
\cite{Skenderis:2002vf}. We will focus on the cases with $m-n=2$. In all these cases, a single scalar field describes the embedding. These probe branes are
thus described by a scalar in a fixed gravitational background, only
with an action more complicated than that of a free scalar. In
general the non-linear Born-Infeld action would lead to new
counterterms. We will show, however, that in the class of embeddings
we analyze the counterterms are identical to that of a free
scalar. To show this it is sufficient to show that the leading 
asymptotic structure of the field equations (i.e. the orders 
where the normalizable and non-normalizable modes appear)
is the same as that of the free scalar, 
and establish that the on-shell actions agree up to
order $r^m$, the highest order at which divergences 
could appear in the action.
From this follows (upon use of (\ref{reg})) that the radial canonical momenta
are the same up to $\pi_{(\D)}$ and thus the counterterm actions
are the same. It also follows (and we will verify it explicitly)
that the solution of the EOM 
has the same form as the free scalar solution to order $r^{\D}$. 

In three of the four cases we consider, the scalar is a coordinate on the
$S^5$ transverse to the brane. The $S^n$ is a trivial cycle on the $S^5$, so the scalar, which
gives the position of the $S^n$ in a transverse direction, can have
a profile such that the $S^n$ ``slips off" the $S^5$ at some finite
$r$. In the one
exceptional case, for the D5 probe, the scalar is not a coordinate
on the $S^5$ but is the position of the $AdS_4$ the brane wraps
inside $AdS_5$. This case is not included in the analysis of this section.

The scalar ``slipping mode'' has mass-squared $M^2 = -n$ (as we will
show). Via the AdS/CFT dictionary the dimension of the dual
operator is $\D = n$. In particular, this means that $d - \D = (m-1)
- n = 1$, so in every case we consider the scalar's asymptotic expansion will
have a leading behavior of simply $r$. Recall also that the
logarithmic term in the asymptotic expansion of the free scalar
entered at order $r^{\D}$ which here will be $r^n$. The cases we
consider are summarized in the following table of useful quantities:

\bigskip

\begin{center}

\begin{tabular}{|c|c|c|c|c|}
\hline
Dp probe & $m = d+1$ & $n = \D$ & $2\D - d$ & -$\frac{d^2}{4}$ \\
\hline
D7 & 5 & 3 & 2 & -4\\
\hline
D5 & 4 & 2 & 1 & -$\frac{9}{2}$\\
\hline
D3 & 3 & 1 & 0 & -1 \\
\hline
\end{tabular}

\end{center}
\bigskip

Given appropriate coordinates on $AdS_5$, the metric of the $AdS_m$
subspace will have the same form as (\ref{metric}) but now $\mu,
\nu = 0, \ldots, m-1$ and $i,j = 0, \ldots, m-2$. Let the $S^5$
metric be \beq d\Omega^{2}_{5} = d\Phi^2 + \sin^{2}(\Phi) d\theta^2
+ \cos^{2}(\Phi) d\Omega^{2}_{3} \eeq with $d\Omega^{2}_{3}$ the
standard metric on $S^{3}$. The $S^n$'s with $n < 3$ sit inside this
$S^3$. The DBI action is then

\beq S = \int d^{m}x \sqrt{G} \cos^{n}(\Phi) \sqrt{1+G^{\m \n}
\partial_{\m}\Phi \partial_{\n}\Phi} \label{DBI} \eeq

\noindent where we neglect integration over the remaining
$n$ variables because the cases we are interested in have no dependence on
these coordinates. The resulting EOM for $\Phi$ is then
\beq 
0 = \Box \Phi + n \tan(\Phi) -\frac{1}{2} \frac{ G^{\m \n}
\partial_{\m} \Phi \partial_{\n} ( G^{\r \s} \partial_{\r} \Phi
\partial_{\s} \Phi  ) }{1 + G^{\a \b} \partial_{\a} \Phi
\partial_{\b} \Phi} \label{DBI-EOM} 
\eeq
where $\Box$ is with respect to the $AdS_m$ metric. To order
$r^{\D}$, this is identical to the EOM of a free scalar with $M^2 =
-n$ in a fixed $AdS_m$ background, which we explicitly show below.
In the case of the free scalar the asymptotic solution has 
two undetermined coefficients, the source and the vev part, 
as we reviewed in section 2. This matches the fact that the 
field equation 
is a second order linear differential equation in $r$, so 
two pieces of boundary data are required in order 
to specify the solution. This is also consistent with holography:
on the field theory side the form of the Lagrangian and the 
specific vacuum we consider (the vev's) are the only information needed
in order to specify the theory. We therefore expect that these
facts will remain true for the asymptotic solution of the 
DBI field equation (\ref{DBI-EOM}). Indeed, despite the higher derivative
terms in the DBI action, the EOM
(\ref{DBI-EOM})
is still a second order differential equation in $r$ (now 
non-linear) and we now verify this expectation explicitly.

Consider a solution with the most general possible form:
\beq 
\Phi(x,r) = r [\sum_{i=0}^{\infty} \phi_{(i)} r^i +
\sum_{j=0}^{\infty} \psi_{(j)} r^j \log(r) + \sum_{k=0}^{\infty}
\sum_{l=2}^{p} \Psi_{(k,l)} r^k \log(r)^l ]. \eeq We
are allowing higher powers of $\log(r)$, but inserting 
this solution into the EOM 
shows that generically truncating the series at any finite order 
$p$ requires the
$\Psi_{(k,l)}$ to all be zero. To see this, consider the leading
term in the expansion of the EOM:
\bea
0 & = & r [ (m-3)\psi_{(0)} + 2\Psi_{(0,2)} + 2 \log(r) ( (m-3)\Psi_{(0,2)} 
+ 3 \Psi_{(0,3)} )) + \ldots \nn
  & + & (p-1) \log(r)^{p-2} ( (m-3) \Psi_{(0,p-1)} + p \Psi_{(0,p)} ) \nn
  & + & p \log(r)^{p-1} (m-3) \Psi_{(0,p)} ]+ O(r^2) \label{logs}
\eea
\noindent Clearly, for any finite $p$, $\Psi_{(0,p)} = 0 $ implies
$\Psi_{(0,p-1)} = 0$ and so on so that all $r \log(r)^l$ do not appear
when $l \geq 2$. A similar story happens for $\Psi_{(1,p)}$, that is, all $r^2 \log(r)^l$ are absent for $l \geq
2$. Of course, an infinite number of higher powers of
$\log(r)$ could appear ($p = \infty$), but this would invalidate the 
assumption that the
solution has a power series expansion. Special cases arise when the numerical coefficients multiplying the highest power of $\log(r)$ are zero. For instance, $m=3$ is a special case in (\ref{logs})
for which the coefficient of $\log(r)^{p-1}$ is zero (although 
in this case one still obtains $\Psi_{(0,p)}=0$).
These cases have to be analyzed separately and could result in higher powers of $\log(r)$ with coefficients determined in terms of the source and the 
vev coefficients. 
For $m=4,5$ we verified that higher powers of $\log(r)$ do not appear up to 
order $r^5$, that is, $\Psi_{(k,l)}=0$ for $k=2,3,4,5$. When $m=3$, $\Psi_{(2,2)}$ and $\Psi_{(2,3)}$ are non-zero but are determined in terms of $\psi_{(0)}$. Notice that 
such higher order logs also appear in the Coulomb branch example in
\cite{Bianchi:2001de} and that both examples with higher powers
of logs are cases where the BF bound
is saturated. It is unclear to us whether this is a general feature
(i.e. that higher order logs appear only when the BF bound is saturated)
or it is specific to the examples we considered.

What results from this discussion is that the form of the solution, up to the 
order we are interested in, is the
same as that of a free scalar:
\beq 
\Phi(x,r) = r [\sum_{i=0}^{\D -1} \phi_{(i)} r^i +
\sum_{j=0}^{\D -1} \psi_{(j)} r^j \log(r) ] + O[r^{\D+1}]. \label{sol} 
\eeq
Inserting this into the EOM gives, to order
$r^2$, \beq 0 = (m-3) \psi_{(0)} r + ( (m-4)\phi_{(1)} + ( m-5 +
(m-4)\log(r) ) \psi_{(1)} ) r^2 + O(r^3) \label{dbicoeffs} \eeq To
this order, this is identical to a free scalar in $AdS_m$ with
mass-squared $M^2 = -n$ and $d - \D = 1$. In particular, the leading
coefficient, $\phi_{(0)}(x)$ is left undetermined by the EOM, as
expected: this will be a source or will contribute to the vev of the
dual operator, depending on whether the scalar saturates the BF
bound. Notice also that for all $m$ that we consider, $\psi_{(1)}$
is zero. For $m=3, 4$, (\ref{dbicoeffs}) determines the form of
solution up to order $r^{\D}$. The value $m=5$, however, requires
the next order, $r^3$. When $m=5$, (\ref{dbicoeffs}) requires $\psi_{(0)} = \phi_{(1)}
= 0$ in which case the $r^3$ term is

\bea 
0 & = & r^3 [ \Box_0 \phi_{(0)} + \tr( g_{(2)})
\phi_{(0)} +\frac{1}{3} (m-5) \phi_{(0)}^3 - 2(m-5) \phi_{(2)} \nn  &
+ & (7-m) \psi_{(2)} + 2(5-m) \log(r) \psi_{(2)} ] + O(r^4). 
\eea
This shows that $\phi_{(2)}$ is undetermined and $\psi_{(2)}$ is
fixed by $\phi_{(0)}$ in precisely the same way as the free scalar, (\ref{psi2}), with $d = m-1 = 4$. This is a nontrivial result:
for all values of $m$ we consider, the form of the solution is the
same as that of a free scalar up to order $r^{\D}$.

Since the solution is the same up to order $r^{\D}$
so will the momentum dilatation 
eigenstates up to $\pi_{(\D)}$. It follows that the 
divergences will be the same as that of a free scalar, 
and we explicitely verify this now.
Expanding the argument of (\ref{DBI}) we obtain:
\bea 
S & = & \int d^{m}x \sqrt{G} [ 1 + \frac{1}{2} [G^{\mu \nu}
\partial_{\mu} \Phi \partial_{\nu} \Phi - n \Phi^2] \nn
  & + & (-\frac{n}{12} + \frac{n^2}{8})\Phi^4 - \frac{1}{8} (G^{\mu \nu}
\partial_{\mu} \Phi \partial_{\nu} \Phi)^2 - \frac{n}{4} \Phi^2 G^{\m \n} \partial_{\m} \Phi \partial_{\n} \Phi + O(r^6) ]. 
\eea
First notice that the leading term is simply the volume of $AdS_m$,
requiring the counterterms $L_1$, $L_2$ and $L_3$ of the last section.

The leading $\Phi$ dependence is that of a free scalar in a fixed $AdS_m$
background with $M^2 = -n$, as advertised. As $\sqrt{G} = \frac{1}{r^m}
\sqrt{g}$ and we care only about $m \leq 5$, all terms of order $r^6$ or
higher cannot contribute to divergences. When $m < 5$ the $\Phi^4$ terms, which go at leading order as $r^4$, will not produce divergences, and hence in these
cases the free scalar counterterms $L_4$ and $L_5$ are sufficient. When
$m=5$ these terms may produce logarithmic divergences, but these terms cancel against each other to order $r^4$ because $n=3$:
\beq
 (-\frac{n}{12} + \frac{n^2}{8})\Phi^4 - \frac{1}{8} (G^{\mu \nu}
\partial_{\mu} \Phi \partial_{\nu} \Phi)^2 - \frac{n}{4} \Phi^2 G^{\m \n} \partial_{\m} \Phi \partial_{\n} \Phi =
 [ (-\frac{n}{12} + \frac{n^2}{8} ) -\frac{1}{8} - \frac{n}{4} ] \phi_{(0)}^4 r^4 + O(r^6)
\eeq

\subsection{Supersymmetric Background Solution}

In the three cases with a scalar describing the slipping mode it is
easy to check that with $g_{ij}(x,r) = \eta_{ij}$ an exact solution 
to the EOM preserving half the supersymmetries is 
\beq \label{background} \Phi = \arcsin(c r) = c r
+ \frac{c^3}{6} r^3 + \ldots, \eeq that is, a solution of the form (\ref{sol}) with $\phi_{(0)} = c$, $\phi_{(1)} = 0$, $\phi_{(2)}= \frac{c^3}{6}$ and no logarithmic terms. This solution describes a D-brane extending up to 
\beq 
r_{max} =\frac{1}{c}. 
\eeq $c$ 
sets the mass of a fundamental matter field in
the dual field theory. In the flat embedding space this embedding
just describes a planar D-brane locate at a fixed distance away from
the stack of D3-branes, \beq X=\frac{1}{r} \sin(\Phi) = c \eeq where
$X$ is one of the flat embedding space coordinates written in a
spherical coordinate system.

We will be mostly interested in calculating correlation functions,
in particular the one-point function, in this particular background.
In order to ensure that holographic renormalization gives finite
answers for all correlation functions we need to make sure however
to add all the counterterms for arbitrary boundary values of the
fields, so that we get finite answers for expectation values in the
presence of sources. While the counterterms and the background
solution are very similar in all the cases, the physical
interpretation as well as the details of the procedure are quite
different in all the cases and hence we will turn to the examples
one by one.

\section{The D3/D7 System} \label{d7}

\subsection{The System}

In the supersymmetric embedding of a probe D7 in $AdS_5 \times S^5$,
the D7 fills the 5D spacetime and wraps an $S^3$ inside the internal
space, that is, the probe wraps an $AdS_5 \times S^3$ and hence
$m=5$, $n=3$, and the scalar's mass is above the BF bound. This system was proposed in \cite{Karch:2002sh} as a
way to introduce flavor, that is fundamental representations, into
the AdS/CFT correspondence. The particular supersymmetric case we
are studying is dual to the ${\cal N}=4$ SYM theory coupled to a
fundamental hypermultiplet that preserves ${\cal N}=2$
supersymmetry. Already in \cite{Karch:2002sh} it was noted that the
gravity side has a non-zero subleading term $\phi_{(2)}$ in its
asymptotic expansion, even though from the field theory side it is
clear that no vev is allowed. Some arguments were given that on the
gravity side the vev really is zero. Starting from
\cite{Kruczenski:2003be} it has become common practice to read off
the vev from the subleading term in the flat embedding space
coordinate $X=\frac{1}{r} \sin(\Phi)$ which on the background
solution is constant and indeed has no subleading term. The absence
of any rigorous derivation of the correct procedure to calculate the
vev became apparent with the work of \cite{Evans:2004ia} where 
different answers were found depending on what coordinate system was
used. In their work a very natural choice was again ``apparent''
that yielded the expected answers, but the question arose how one
would find the right coordinate system in less symmetric cases. Here
we will apply holographic renormalization to that problem and will
find an unambiguous answer.

\subsection{Counterterms}

In addition to the counterterms we derived for general $m$
and $n$ above, only in the case of the D7 brane we face an
interesting new subtlety: the possibility of finite counterterms. A
counterterm of the form \beq L_f = \alpha \sqrt{\gamma}
\Phi(x,\epsilon)^4\eeq will not introduce divergences for any
$\alpha$, but will change the on-shell action by a finite amount
depending on the free coefficient $\alpha$. Different values of
$\alpha$ correspond to different renormalization schemes and some correlation functions will contain scheme-dependent terms. As
first pointed out in \cite{Bianchi:2001de} in the special case that
one is interested in correlation functions in a {\it supersymmetric}
background, one can fix the finite counterterms by picking the
unique scheme in which supersymmetry is preserved. In the
supersymmetric scheme not just the divergent pieces of the on-shell
action but the action as a whole has to vanish when calculated on
the supersymmetric background solution. Otherwise the ground state
energy in the dual field theory would not be zero and hence
supersymmetry would be broken.

For the D7 case the counterterms read
\bea
L_{1} &=& - \frac{1}{4} \sqrt{\gamma} \\
L_2 &=& - \frac{1}{48} \sqrt{\gamma} R_{\gamma} \nn
L_3 &=& - \log \epsilon \sqrt{\gamma} \frac{1}{32} ( R_{ij} R^{ij} -
\frac{1}{3} R_{\gamma}^2 ) \nn
L_4 &=& \frac{1}{2} \sqrt{\gamma} \Phi^2(x,\epsilon) \nn
L_{5} &=& -\frac{1}{2} \log(\epsilon) \sqrt{\gamma} \Phi(x,\e)
(\Box_{\gamma} + \frac{1}{6}  R_{\gamma}) 
\Phi(x,\e) \nonumber
\eea

For the supersymmetric embedding, $g_{ij}(x,r) = \eta_{ij}$ and the curvature on the slice is zero so only $L_1$ and $L_4$ contribute. To fix the finite counterterm we only need to plug the background
solution (\ref{background}) into the action. On the
background solution the action exactly reduces to \beq S_{reg}= \int d^4 x
\int_{\epsilon}^{r_{max}} dr \frac{1}{r^5} ( 1 - c^2 r^2 ) =\int d^4
x \left ( \left . -\frac{1}{4} \frac{1}{r^4} + \frac{1}{2}
\frac{c^2}{r^2} \right |_{\epsilon}^{r_{max}} \right ) . \eeq Indeed
the action picks up a non-vanishing contribution of $-\frac{1}{4
r_{max}^4} + \frac{c^2}{r_{\max}^2} = +\frac{c^4}{4} $ from the IR
boundary. The UV divergence is ensured to cancel by the counterterms
we introduced above. $L_4$ gives an additional finite piece, $L_4 = \ldots + \frac{c^4}{6} + \ldots$.
All in all we see that in order to set the on-shell action to zero we need \beq \alpha=- \frac{5}{12}. \eeq

\subsection{One-point function}

Holographic renormalization tells us that the vev is obtained as the
$\epsilon \rightarrow 0$ limit of
\beq 
\langle O \rangle = \frac{1}{\epsilon^3 \sqrt{\gamma}}
\frac{\delta S_{sub} }{\delta \Phi(\epsilon)}. 
\eeq 
In this case,
\beq 
\frac{\delta S_{reg}}{\delta \Phi} = \left . \frac{ \delta
L}{\delta \Phi'} \right |_{\epsilon} = -
\frac{1}{\epsilon^3} \sqrt{g} \cos^{3}(\Phi(\epsilon))
\frac{\Phi'(\epsilon)}{\sqrt{ 1 + \epsilon^2 \Phi'(\epsilon)^2}}.
\eeq 
This then contributes divergent and finite parts to the vev
\beq 
\langle O \rangle_{reg} = - \frac{\phi_{(0)}}{\epsilon^2} -3 \psi_{(2)}
\log(\epsilon) + (2 \phi_{(0)}^3 - 3\phi_{(2)} - \psi_{(2)}) + O(\epsilon^2)
\eeq
Adding in addition the variations of the counterterms we finally
obtain for the vev
\beq \langle O \rangle = \log(\epsilon) (-2\psi_{(2)} -\frac{1}{6}
\phi_{(0)} R_0 - \Box_0 \phi_{(0)}) + (\frac{1}{3} \phi_{(0)}^{3} 
- 2\phi_{(2)} - \psi_{(2)}) + O(\epsilon^2). \eeq
The coefficient of the $\log(\epsilon)$ is zero via the EOM,
(\ref{psi2}), and inserting the value of $\psi_{(2)}$ from
(\ref{psi2}) into the finite piece then gives
\beq \langle O \rangle = -2\phi_{(2)} + \frac{\phi_{(0)}^{3}}{3} +
\frac{R_0}{12} \phi_{(0)} + \Box_0 \phi_{(0)}. 
\eeq
The last two terms come from the curvature of the induced  metric on
the $r = \epsilon$ slice and can in fact be eliminated by adding a
finite counterterm proportional to the matter conformal anomaly. On
the background solution $\sin(\Phi) = c r$ we have $\phi_{(0)}=c$ and
$\phi_{(2)}=\frac{c^3}{6}$, so indeed \beq \langle O
\rangle_{background} = 0. \eeq The finite counterterms required by
supersymmetry set the vev to zero despite the appearance of a
non-vanishing subleading term in the asymptotic expansion of the
scalar field.

\section{The D3/D5 System}
\label{d5}

\subsection{The System}

In the supersymmetric embedding of a probe D5 in $AdS_5 \times S^5$,
the D5 fills an AdS$_4$ subspace of the 5D spacetime and wraps an
$S^2$ inside the internal space, that is, the probe wraps an $AdS_4
\times S^2$ and hence $m=4$, $n=2$ and the scalar's mass is above the BF bound. This system was introduced in
\cite{Karch:2000gx} and analyzed in detail in \cite{DeWolfe:2001pq}.
The dual field theory in this case is a defect conformal field
theory. Again the probe brane in the bulk corresponds to the
introduction of fundamental matter in the field theory, but this
time the matter only lives on a 3d defect in the 4d field theory.
While the defect breaks translation invariance in the directions
orthogonal to the defect, the field theory remains conformal as long
as the matter is massless. The subgroup of the full conformal group
that is preserved is the $SO(3,2)$ that leaves the position of the
defect invariant. This symmetry in the bulk demands the worldvolume
to be AdS$_4$. The supersymmetric background solution once more
describes adding a mass term to the fundamental matter. A new
possibility that arises in this case is to turn on the scalar that
corresponds to the embedding of the AdS$_4$ inside the AdS$_5$
\cite{Skenderis:2002vf}. This deformation breaks all supersymmetries. 
On the field theory side it was argued in \cite{Skenderis:2002vf} that 
one turned on a vacuum expectation value for the defect field, as we will
verify explicitly below. We will also see that the energy of the 
configuration in non-zero. On the field theory side one certainly
expects such configurations to eventually relax to the supersymmetric
ground state. On the gravity side,  this process should be mapped to some
instability of the brane embedding that drives the vev to zero.
It would be interesting to analyze the stability of this embedding.
This would provide
an example of an unstable (or perhaps metastable) configurations that 
exists and can be studied both at weak and strong coupling.

\subsection{Mass deformation}

In order to cancel the UV divergences we need to add the following
counterterms:
\bea
L_{1} &=& - \frac{1}{3} \sqrt{\gamma} \\
L_2 &=& - \frac{1}{12} \sqrt{\gamma} R_{\gamma} \nn
L_4 &=& \frac{1}{2} \sqrt{\gamma} \Phi^2(x,\epsilon) \nn
 L_{5} &=&   
-\frac{1}{2} \sqrt{\gamma} \Phi(x,\e) (\Box_{\gamma} + \frac{1}{4}R_{\gamma})
\Phi(x,\e) \nonumber
\eea 
No option to add finite counterterms involving only
$\Phi$ and $\sqrt{\gamma}$ arises. Indeed with this set of
counterterms the on-shell action on the background solution is zero
already, 
\beq 
S_{reg}= \int d^3x \int_{\epsilon}^{r_{max}} dr \frac{1}{r^4}
\sqrt{ 1 - c^2r^2 } = \int d^3x \left ( \left . - \frac{(1-c^2
r^2)^{3/2}}{3 r^3} \right |_{\epsilon}^{r_{max}} \right ). 
\eeq 
The IR term, with $r_{max} = \frac{1}{c}$, is identically zero.

With these counterterms it is again straightforward to calculate the vev. Setting to zero the terms involving $R_{\g}$ and $\Box_{\g}$ one simply obtains 
\beq 
\langle O \rangle= - \phi_{(1)}.
\eeq
Again, the supersymmetric background
solution which has $\phi_{(1)}=0$ has no vev, as expected.

\subsection{Vev deformation}

In this subsection we study the probe RG flow obtained 
by turning on a scalar that controls the
embedding of the AdS$_4$ part inside the AdS$_5$ while keeping the
brane wrapped on the maximal sphere inside $S^5$, $\Psi=0$ 
\cite{Skenderis:2002vf}. For simplicity in this case we
will only calculate the vev on the background solution; no problems
are anticipated in once more working out the full holographic RG
program for arbitrary correlation functions. Looking for an
embedding $x_3=x(r)$ the induced metric on the 4d part becomes 
\beq
ds^2 = \frac{1}{r^2} (-dt^2 + dx_1^2 + dx_2^2 + (1 + (x')^2) dr^2 )
\eeq 
and the corresponding action is 
\beq 
S = \int \frac{d^4x}{r^4} \sqrt{1 + (x')^2}. \eeq
The most general solution to the equations of motion with the ansatz
that $x_3$ only depends on $r$  is 
\beq 
x' = \frac{c r^4}{\sqrt{1 -c^2 r^8}} \, ,
\eeq 
so that asymptotically 
\beq 
x = r_0 + \frac{c}{5} r^5 + \ldots \, . 
\eeq
This brane extends up to
\beq
r_{max}=\frac{1}{c^{1/4}}\, .
\eeq
The solution with $c=0$ is the supersymmetric AdS$_4$ $\times$ $S^2$
defect. What happens at $r=r_{max}$ for the vev deformation is quite distinct
from what we have encountered in the other examples. While before a sphere
was shrinking in the internal space so that the brane smoothly terminates,
for the vev deformation the internal sphere has constant size. To
understand the physics of this embedding it is useful to note that
close to $r_{max}$ we simply get
\beq
\label{closetormax} x' = \frac{\sqrt{r_{max}}}{2 \sqrt{2}} 
\frac{1}{\sqrt{r_{max} - r}} + \ldots, 
\,\,\,\,\,\,\,\,
\,\,\,\,\,\,\,\,
x = \sqrt{\frac{r_{max}}{2}} \sqrt{r_{max} - r} + \ldots .
\eeq
Instead of ending at $r=r_{max}$ the string turns around and hits the 
boundary again. The probe brane no longer is dual to the defect conformal
field theory of a single D5 brane defect, but with a D5 and anti-D5 defect
seperated by a distance $d = 2 \frac{ \sqrt{\pi} \Gamma(13/8)}{5 \Gamma(9/8)}
r_{max}$.

The on-shell action can be evaluated exactly 
\beq \label{ons}
S_{reg} = \int d^3 x \int_{\epsilon}^{r_{max}} dr
\frac{1}{r^4 (1-c^2 r^2)} = \int d^3 x 
\left(\frac{1}{3 \epsilon^3} - c^{3/4} 
\frac{\sqrt{\pi} \G(13/8)}{15 \G(9/8)} + O(\e)\right)
\eeq 
where the finite term comes from the IR boundary. 
It follows that we should add to the action the 
counterterm
\beq L_1 = - \frac{1}{3} \sqrt{\gamma}\, . 
\eeq 
With this counterterms the action is not just finite but zero
for the supersymmetric embedding, and is negative 
for the non-supersymmetric embedding. For the general $x(r)$
solution we get for the vev
\beq 
\langle O \rangle = \frac{1}{\epsilon^3}
\frac{1}{\sqrt{\gamma}} \frac{\delta S_{sub}} {\delta x} =
\left . -\frac{1}{\e^4} \frac{x'}{\sqrt{1 + (x')^2}} \right
|_{\epsilon} = - c. 
\eeq
This agrees with the expectation that a non-vanishing $c$
corresponds to a vev-induced flow once one realizes that
$\frac{x}{r}$ is dual to an operator of dimension four
\cite{DeWolfe:2001pq}. The negative energy does not signal
an instability of the original D5 brane defect since, as we noted above,
the vev deformation really is dual to a system with 2 defects, one
due to a D5 and one due to an anti-D5. It is not too surprising that latter
system can lower its energy by moving out onto the Higgs branch. Since
the constant $c$ is completely fixed by the distance $d$ between the 
defects we expect this configuration to correspond to the stable ground
state of the system.

\section{The D3/D3 System} \label{d3}

\subsection{The System}

The D3/D3 system is another example that can be treated with the
same methods. In this case, the probe D3's wrap an $AdS_3  \times
S^1$ inside $AdS_5 \times S^5$, $m=3$ and $n=1$. This system was analyzed in \cite{Constable:2002xt}. 
It corresponds once more
to a defect conformal field theory, this time with the fundamental
hypermultiplet localized on a 2d defect in the 4d field theory. The
interesting new aspect of this system is that the mass of the
slipping mode saturates the BF bound. This gives rise
to a slightly different structure of counterterms in holographic
renormalization; it also leads to an interesting reversal in the
role of the vev versus mass as we will see below.

From the boundary theory point of view it is also clear that this
system is different. In all the other cases we had a higher
dimensional brane intersecting the D3 brane, so that in the field theory
limit the worldvolume fields on the higher dimensional brane
decoupled together with gravity. The only dynamical fields remaining
were those on the D3 branes and on the intersection from the 3-7 or
3-5 strings respectively. For the D3/D3 system the ${\cal N}=4$ SYM
on both D3 stacks remains dynamical, so the field theory is really
the theory of two ${\cal N}=4$ theories coupled to each other via a
bifundamental hypermultiplet living on a 2d defect. In particular,
the spacetime separation between the two stacks of D3 branes can be
viewed as a vev for one of the additional D3's worldvolume fields
and should no longer be interpreted as a parameter (as in the D5
or D7) case, but as a vev of a dynamical field. We will see that
holographic renormalization correctly accounts for this effect. It
is precisely separating the two stacks in spacetime that is
described by the supersymmetric background solution we studied for
general $m$ and $n$, and we will see that unlike in the other cases
for $m=3$, $n=1$ it is a vev instead of a mass deformation.
Implementing holographic renormalization properly we also calculate
the two-point function in this case, yielding a different
answer from the one originally advertised in
\cite{Constable:2002xt}.

\subsection{Counterterms and One-Point Function}

The interesting change in roles between source and vev can already
be observed from the form of the asymptotic solution which in this
case reads \beq \Phi = r (\tilde{\psi}_{(0)} + \psi_{(0)} \log(r) + \ldots
)\eeq for which the first term is normalizable and will set the vev,
while the second term is non-normalizable and will correspond to a
source in the dual theory. On our supersymmetric background
solution the source term is zero; we will calculate the
vev momentarily.
For $m=3$, $n=1$ the counterterms become 
\bea
L_{1} &=& - \frac{1}{2} \sqrt{\gamma} \\
L_2 &=& \frac{1}{4} \log \e \sqrt{\gamma} R_{\gamma} \\
L_4 &=& \frac{1}{2} \sqrt{\gamma} \Phi^2(x,\epsilon) ( 1 + \frac{1}{\log \e} ) \eea
No finite counterterms can arise and it is again easy to verify that
the on-shell action for our supersymmetric background vanishes with
these counterterms already. 

In this case, the vev is the $\epsilon \rightarrow 0$ limit of \beq
\langle O \rangle = \frac{\log(\epsilon)}{\epsilon}
\frac{1}{\sqrt{\gamma}} \frac{\delta S_{sub} }{\delta
\Phi(\epsilon)}. \eeq
Using the counterterms it is again straightforward to obtain 
\beq \langle O \rangle = \tilde{\psi}_{(0)} 
\eeq
where we again set to zero the terms involving the curvature on the
slice. The vev is thus indeed determined by the coefficient of the
normalizable mode; on our supersymmetric background we simply have $\langle O \rangle = c$. As expected, the brane separation shows up as a vev and no longer as a mass term.

\subsection{Two-point functions}

We discuss in this section the computation of correlation functions. 
$n$-point functions
can be obtained from the exact one-point function  by further
differentiating $(n-1)$ times w.r.t. the source. It follows that in
order to obtain the two-point function it is sufficient to solve the
linearized equation of motion. 

The simplest case to consider is the two-point function for the theory
specified by the embedding $\Phi=0$.  We first discuss the 2-point function
of the operator dual to the scalar field $\Phi$. This can be
obtained by linearizing the field equation
(\ref{DBI-EOM}) around  $\Phi=0$,
and Fourier transforming in the spatial directions we get
\beq 
r^2 \Phi''- r \Phi' -(k^2 r^2 -1) \Phi=0 
\eeq
This is equal (as it should be) to the linearization of the field
equations of a scalar of mass $M^2=-1$. The solution of this
equation that is regular in the interior is
\beq \Phi(r,k) = r K_0(k r). \eeq
Expanding for small $r$ we get
\beq \Phi(r,k)=\psi_{(0)}(k) r \left(\log r + \frac{1}{2} \log
\frac{k^2}{\mu^2} \right) \eeq
where $\psi_{(0)}(k)$ represents the overall normalization of the
solution, and $\mu^2 = 4 e^{-2 \gamma}$ ($\gamma$ is the Euler
constant). Proper units are restored using the AdS radius. It
follows that
\beq
\tilde{\psi}_{(0)}(k) =   \psi_{(0)}(k) \frac{1}{2} \log \frac{k^2}{\mu^2}
\eeq
\beq
\langle O(k) O(-k)\rangle = -
\frac{\delta \tilde{\psi}_{(0)}}{\delta \psi_{(0)}} = -
\frac{1}{2} \log \frac{k^2}{\mu^2}
\eeq
Fourier transforming we get the renormalized
version of $1/x^2$ (see appendix)
\beq
\langle O(x) O(0)\rangle = \frac{1}{2 \pi} \left(\frac{1}{x^2}\right)_R
\eeq
(with $m=1$),
as is appropriate for the two-point function for a scalar operator 
of dimension one.

We now discuss the computation of 2-point function which are dual to
fields parametrizing the flunctuations of $AdS_3$ in $AdS_5$. The mass
of the fluctuations is \cite{Constable:2002xt},
\beq
m^2 = (l-1) (l-3)
\eeq
with $l$ a positive integer. One 
should distiguish between the fluctuations with mass
\beq
m^2 > -\frac{d^2}{4}+1=0  \qquad \Rightarrow \qquad l>3
\eeq
in which case there is a single branch of dual operators with 
dimension,
\beq \label{+br}
\D=l-1, \qquad l>3
\eeq
and fluctuations with mass
\beq
-\frac{d^2}{4}=-1 \leq m^2  \leq -\frac{d^2}{4}+1 =0, \qquad 
\Rightarrow \qquad 1 \leq l \leq 3
\eeq
In this case there are two inequivalent quantizations and 
we have two branches of operators. The $l=2$ correspond to
$m^2=-1$, so the fluctuations saturate the BF bound. We discussed
the computation of the 2-point functions for this case in the beginning 
of this section. The $l=1$ and $l=3$ cases correspond to massless fluctuations.
The $l=1$ case is special: the corresponding boundary operator has 
dimension zero and thus saturates the unitarity bound.
From the bulk point of view this case utilizes the $\D_-$ branch and 
the roles of source and  vev are exchanged. The computation of correlation 
functions for such cases has been discussed in  \cite{Klebanov:1999tb}.
When the operators saturate the unitarity bound, however,
the normalization of the (holographically computed) 
2-point function turns out to vanish (set $\D=d/2-1$ in (2.19) or (2.21) 
of \cite{Klebanov:1999tb}) even though the corresponding
QFT 2-point function is generically non-zero indicating subtleties
with the normalization of these operators. Leaving this
special case aside we now discuss the computation of 2-point functions
for $l \geq 3$.

In our analysis so far we have constructed the counterterms 
required to render the on-shell action finite and then 
derived the 1-point function. This can also be done for the 
case at hand and this would {\it uniquely} fix (up to finite
scheme dependent terms) all boundary terms. We should also
note that another (perhaps more fundamental) requirement
on the action is that it is stationary when the field
equations hold and appropriate boundary conditions are imposed. 
This also fixes all boundary terms
to be the ones determined by finiteness \cite{Papadimitriou:2005ii}.

In this section we are only interested in computing 2-point functions, 
however, so instead of first computing the counterterms that would 
render finite any correlation function
we  will follow the simpler route developed in \cite{PS2}.
2-point functions are determined by solving the linearized 
fluctuation equations and to renormalize we need
to perform a near-boundary analysis of the fluctuation equations.

The fluctuation equations can be obtained from
(3.16) of \cite{Constable:2002xt} \footnote{We employ the $\tilde{w}$ notation of \cite{Constable:2002xt}, whereas \cite{Arean:2006pk} uses $w$. The two are related by $w = r \tilde{w}$}. In our coordinates these read
\beq \label{fl_l}
r^2 \tilde{w}_l''- r \tilde{w}_l' -(k^2 r^2 +m^2) \tilde{w}_l=0 
\eeq
where we Fourier transformed along the boundary directions.
In the recent work \cite{Arean:2006pk} the fluctuation 
equations were analyzed for D3 branes separated by a distance
$L$ and shown to be of the hypergeometric type. One can thus
readily derive the 2-point functions in this more general setting
but for simplicity we restrict to the $L=0$ case. The fluctuation
equation in (\ref{fl_l}) can also be obtained from (4.10) of 
\cite{Arean:2006pk} in the limit $L \to 0$. The solution of
(\ref{fl_l}) that is regular in the interior is
\beq
\tilde{w}_l(r,k)=r K_{(l-2)} (k r)
\eeq

We now discuss how to extract {\it renormalized} 2-point functions
from the solution of the fluctuation equation.
We will follow the Hamiltonian formalism described in the introduction.
It is useful to introduce a new radial coordinate
$\r =- \log r$. In these coordinates the $AdS_3$ metric reads
\beq
ds^2 = d \r^2 + e^{2 \r} d x^i dx^i
\eeq
and the canonical momenta are given by\footnote{Strictly speaking, the
momenta are densities and one should include a factor of $\sqrt{\g}$ 
as in the formulae in the introduction. The equations are simpler without 
these factors, however, so we will work with (\ref{mom_def}).}
\beq \label{mom_def}
\pi_l = \dot{\tilde{w}}_l
\eeq
where overdot denotes differentiation w.r.t. $\r$.  The  
fluctuation equation (\ref{fl_l}) becomes
\beq \label{mom}
\dot{\p}_l + 2 \p_l - (\xi + m^2) \tilde{w}_l=0
\eeq
where we defined $\xi = e^{- 2 \r} k^2$, which comes from the Fourier 
transform of $\Box_\g$ at the regulating surface $r=const$ and 
$\g_{ij}=e^{2 \r} \d_{ij}$ is the corresponding induced metric.

Since we are interested in 2-point functions, we need to know
the radial canonical momentum as a function of the induced field to linear 
order. Covariance fixes their relation to be the form
\beq \label{pi_2}
\pi_l = f_l(\xi) \tilde{w}_l.
\eeq 
The {\it regularized} 2-point function (with the regulator being
a small constant value of the radial coordinate $r=\e$)
is then given by
\bea
\langle O_{l-1}(k) O_{l-1}(-k)\rangle_\e &=& 
- \frac{1}{\e^{2(l-2)}}f_l(\xi) \nonumber \\
&=& -\frac{1}{\e^{2 (l-2)}}
\left((l-3) + k \e \frac{K_{(l-3)}(k \e)}{ K_{(l-2)}(k \e)}\right)
\eea
In the limit $\e \to 0$ these correlators diverge. Our 
task is to extract correctly the finite part.

As reviewed in section \ref{holorg}, the {\it renormalized} 1-point
function in the presence of sources is given by the part of the canonical momentum with dilatation
weight equal to the dimension of the dual operator 
\cite{Papadimitriou:2004ap,PS2}. To extract this part we
need the asymptotic form of $f_l(\xi)$. To this end,
we insert (\ref{pi_2}) in (\ref{mom}) to obtain
\beq \label{f_eqn}
\dot{f}_l + f_l^2 + 2 f_l - (\xi + m^2) =0
\eeq
From the general discussion in the introduction
we know that the $f_l(\xi)$ has an expansion of the 
form,
\beq \label{f_exp}
f_l(\xi)=f_{l (0)} + f_{l (2)} + \cdots + f_{l(2 l-4)} + 
\tilde{f}_{l(2 l-4)} \log r^2 + \cdots
\eeq
Furthermore, in our case the dilatation operator in (\ref{dil}) is just equal 
to the radial derivative, 
\beq
\d_D=\partial_\r
\eeq
This follows from the fact that the background is $AdS_3$  
(insert $\gamma_{ij} = e^{2 \r} \d_{ij}, \Phi=0$ in  (\ref{dil})).

Inserting the expansion (\ref{f_exp}) into (\ref{f_eqn}) and organizing the 
terms according to their dimension one obtains all coefficients
except for $f_{l (2l-4)}$ which is left undetermined. Let us 
explain this computation. In (\ref{f_eqn}) the mass term
has dimension 0 and $\xi$ has dimension 2, so $f_{l(0)}$
and $f_{l(2)}$ are different from the remaining $f_{l(2p)}$.
We get
\bea \label{f02}
&&l \neq 3: \qquad f_{l(0)}=l-3, \qquad f_{l(2)} = \frac{\xi}{2(l-3)}  \\
&& l=3: \qquad f_{3(2)}\ {\rm undetermined}, \qquad
 \tilde{f}_{3(2)} = -\frac{\xi}{2} \nonumber
\eea
For the remaining coefficients
one gets a recursive relation,
\bea \label{f2m}
&&p \neq l-2: \qquad f_{l(2p)} = \frac{1}{2 (p-l+2)} \sum_{n=1}^{p-1} f_{l(2(p-n))} f_{l(2n)} \\
&& p=l-2: \qquad f_{l(2 l-4)}\ {\rm undetermined}, \qquad
\tilde{f}_{l(2 l-4)} = \frac{1}{2} \sum_{n=1}^{l-3} f_{l(2(l-2-n))} f_{l(2n)}
\nonumber
\eea

Having determined the asymptotic expansion we now 
extract the renormalized 2-point function as
\beq \label{2pt_ren}
\langle O_{l-1}(k) O_{l-1}(-k)\rangle 
= -\lim_{\e \to 0} \frac{1}{\e^{2 (l-2)}}
\left((l-3) + k \e \frac{K_{(l-3)}(k \e)}{ K_{(l-2)}(k \e)}
-\sum_{p=0}^{l-3} f_{l(2p)} - \tilde{f}_{l(2l-4)} \log \e^2 \right)
\eeq
Let us work out explicitly the first few cases ($\mu^2=4 e^{-2 \gamma}$ 
as before):
\bea
l=3: && \langle O_{2}(k) O_{2}(-k)\rangle 
=\frac{1}{2} k^2 \log \frac{k^2}{\mu^2} \\
l=4: && \langle O_{3}(k) O_{3}(-k)\rangle 
=-\frac{1}{8} (k^2)^2 \log \frac{k^2}{\mu^2}  \nonumber \\
l=5: && \langle O_{4}(k) O_{4}(-k)\rangle 
=\frac{1}{128} (k^2)^3 (\log \frac{k^2}{\mu^2} -\frac{1}{2}) \nonumber
\eea  
which is the correct behavior for two-point operator of dimension 2, 
3 and 4, respectively. (The last term in the r.h.s. of the $l=5$ 
case is scheme dependent. It contributes a contact term in $x$ space). 
Fourier transforming to $x$ space 
(using the results of the appendix) we find
\beq
\langle O_{\D}(x) O_{\D}(0)\rangle = 
\frac{2 \nu \G(\frac{d}{2}+\nu)}{\pi^{d/2} \G(\n)} 
\left(\frac{1}{x^{2 \D}}\right)_R
\eeq
where the subscript $R$ indicates that the expression is 
renormalized and $\nu=\D-d/2$. The normalization of the 2-point functions 
coincides with the one determined in \cite{Freedman:1998tz} (as it should).

By construction, the subtractions in (\ref{2pt_ren}) can be implemented 
by means of counterterms. The explicit form follows
from (\ref{ac_reg}) and is given by
\bea
S_{ct} &=& \frac{1}{4} \sum_{l \geq 3} 
\int d^2 x \sqrt{\gamma} 
\tilde{w}_l \left(\sum_{p=0}^{l-3} f_{l(2p)} + \log \e^2 \tilde{f}_{l(2 l-4)}
\right) \tilde{w}_l^*
+ c.c. \nonumber \\
&=&\frac{1}{4} \sum_{l \geq 3} 
\int d^2 x \sqrt{\gamma} 
\left((l-3) \tilde{w}_l \tilde{w}_l^* 
-\frac{1}{2 (l-3)} \tilde{w}_l \Box_\g \tilde{w}_l^* 
+ \cdots \right) + c.c.\, , \label{count}
\eea
(when $l=3$ there is only the logarithmic counterterm). 
The renormalized action is given by the sum of the bulk action 
and this boundary term. The action derived in \cite{Constable:2002xt}
should be viewed as a ``bare'' action. Adding to it appropriate 
boundary terms leads to a renormalized action and the corresponding
renormalized correlators exhibit power law behavior.
 
The fields discussed here are dual to supersymmetric 
operators that saturate a BPS bound \cite{Constable:2002xt}. 
A realization of these
operators in terms of defect fields has been proposed in 
\cite{Constable:2002xt} and they contain massless scalar 
fields. As discussed there this implies
that in general there will be strong infrared effects. 
The holographic computation discussed here and the fact that
these operators 
are supersymmetric suggest that appropriate 
IR renormalization would lead to 2-point functions
exhibiting power law behavior at weak coupling. We  leave the
investigation of this issue for future work.

\section*{Appendix: Fourier Transform} \label{appendix}

In differential regularization \cite{Freedman:1991tk},
$1/x^2$ is represented by
\beq
\left(\frac{1}{x^2}\right)_R = \frac{1}{8} \Box \log^2 x^2 m^2
\eeq
where the subscript $R$ indicates that this is a renormalized 
expression and $m^2$ is the renormalization scale. The renormalized
expression differs from $1/x^2$ by the infinite 
term $\delta(x) \log x^2 m^2$ localized at $x=0$.

We now compute
the Fourier transform following \cite{Avdeev:1992jp}.
Integrating by parts once and after some manipulations
one arrives at
\bea
I(k) &=& \int d^2 x e^{i k x}  \frac{1}{8} \Box \log^2 x^2 m^2
=- i \frac{k_\mu}{2} \int d^2 x e^{i k x} \frac{x^\mu}{x^2} \log x^2 m^2
\nn 
&=& -\pi
\frac{d}{d a} \left(\left(\frac{k^2}{\mu^2}\right)^a F(a)\right)
\Big |_{a=0}
\eea
where $\mu^2 = 4 e^{-2 \gamma} m^2$ and
\beq
F(a) = -a \frac{2^{2 a+1}}{2 \pi} e^{-2 a \gamma} \int r dr d \theta
\frac{e^{i r \cos \theta}}{r^{2 a +2}}
\eeq
The integral can be readily computed,
\beq
F(a) = e^{-2 \gamma a} \frac{\Gamma(1-a)}{\Gamma(1+a)} = 1 + O[a]^3
\eeq
It follows
\beq
I(k) = - \pi \log \frac{k^2}{\mu^2}.
\eeq

\acknowledgments

A.K. and A.O'B. would like to thank C.R. Graham for helpful
conversations. The work of A.K. was supported in part by DOE
contract \# DE-FG02-96-ER40956. The work of A.O'B. was supported by
a Jack Kent Cooke Foundation scholarship. KS is supported by NWO.

\end{document}